\documentclass[prb,superscriptaddress,twocolumn,preprintnumbers,a4,showpacs,floatfix]{revtex4-1}
\usepackage[T1]{fontenc}
\usepackage[english]{babel}

\usepackage[sort&compress]{natbib}

\usepackage{color}

\usepackage[latin1]{inputenc} % tämän pitäisi toimia oletuksena ainakin Texnic-centerissä ja TeXMakerissa
\usepackage{mathtools}
\usepackage{amsmath}
\usepackage{amsfonts}
\usepackage{amssymb} % Tarvitaan mm. lukualueiden symbolien kirjoittamiseen komennolla \mathbb{}.
\usepackage{amsthm} % mm. tarjoaa lause- ja todistusympäristöt.
\usepackage{graphicx} % Kuvien lisäämistä varten
\usepackage{verbatim} % Tarjoaa mm. kommenttiympäristön.
\usepackage{enumerate} % Lisää vaihtoehtoja numeroitujen listojen muotoiluun.
\usepackage{setspace} % Rivivälin helppoa muuttamista varten.
\usepackage{booktabs} % Tarjoaa \toprule-, \midrule- ja \bottomrule-komennot taulukkoja varten

\newcommand{\be}{\begin{equation}}
\newcommand{\ee}{\end{equation}}

\def\bear{\begin{eqnarray}}
\def\eear{\end{eqnarray}}

\newcommand{\cmtr}[1]{#1}

\begin{document}

%Muutetaan riviväliksi 1,5 tiettyjä paikkoja, kuten kuvatekstejä ja alaviitteitä, lukuun ottamatta.
%\onehalfspacing
\setcounter{page}{1}

\title{Peeling of multilayer graphene generates complex interlayer sliding patterns}

\author{Topi Korhonen}
\author{Pekka Koskinen}
\email[email:]{pekka.koskinen@iki.fi}
\address{NanoScience Center, Department of Physics, University of Jyvaskyla, 40014 Jyv\"askyl\"a, Finland}

\pacs{61.46.-w,62.25.-g,68.65.Pq,68.55.-a}
% 68.65.Pq 	Graphene films
% 62.25.-g 	Mechanical properties of nanoscale systems
%68.35.Gy 	Mechanical properties; surface strains
%68.55.-a 	Thin film structure and morphology
%68.60.-p 	Physical properties of thin films, nonelectronic
%68.60.Bs 	Mechanical and acoustical properties 

%68.65.-k 	Low-dimensional, mesoscopic, nanoscale and other related systems: structure and nonelectronic properties
%61.46.-w 	Structure of nanoscale materials

\begin{abstract}

Peeling, shearing, and sliding are important mechanical phenomena in van der Waals solids. However, theoretically they have been studied mostly using minimal periodic cells and in the context of accurate quantum simulations. Here, we investigate the peeling of large-scale multilayer graphene stacks with varying thicknesses, stackings, and peeling directions by using classical molecular dynamics simulations with a registry-dependent interlayer potential. Simulations show that, while at large scale the peeling proceeds smoothly, at small scale the registry shifts and sliding patterns of the layers are unexpectedly intricate and depend both on the initial stacking and on the peeling direction. These observations indicate that peeling and concomitant kink formations may well transform stacking order and thereby profoundly influence the electronic structures of such multilayer solids.

\end{abstract}
\maketitle

\section{Introduction}

Graphene, carbon nanotubes and graphitic systems have attracted interest because of their unique and appealing electrical and mechanical properties. The proposed range of applications is vast, including the ability to study quantum electrodynamics on bench-top experiments \cite{QEDNovoselov2005}. Graphite, or multilayer graphene, is a van der Waals solid, which means that the interlayer interactions are relatively weak and layers can easily slide with respect to each other and even get peeled. The peeling of graphene layers is also central in the Scotch tape experiments that ultimately led to the discovery of graphene \cite{Novoselov2004}. Due to the easy sliding and peeling, graphite has also long been used as the 'lead' in pencils. 

When multilayer graphene stacks are peeled, they also get bent. However, in multilayer graphene the in-plane Young's modulus is three orders of magnitude larger than the interlayer shear modulus \cite{ElasticConstantDFTTeor}, making the bending properties of multilayer graphene dominated by the interlayer shear. Upon bending the layers within the stacks prefer rather to slide than to compress, similarly to a bent stack of paper. The bending-induced sliding of the graphene layers consequently affects the stacking, which is known to influence profoundly the electronic properties of multilayer graphene \cite{CastroNeto}. Interlayer shifts and registry changes are known to have significant effects also in other materials \cite{Mos}, thus making the studies of peeling and the related bending-induced sliding patterns of multilayer graphene particularly relevant. 

The bending of multilayer graphene stacks have been investigated both experimentally \cite{BendingAndInterlayerShearModulusExp, GraphiteDrumsExper} and theoretically \cite{BendingThinGrapheneTeor, InterlayerShearMultilayerResonatorsTeor, InterlayerShearTeor, ElasticConstantDFTTeor}. However, these studies have mostly focused on the interlayer shear modulus \cite{ElasticConstantDFTTeor, BendingAndInterlayerShearModulusExp} and bending modulus \cite{BendingThinGrapheneTeor, GraphiteDrumsExper} of multilayer graphene stacks. Thus, surprisingly little effort has been invested on investigating the detailed mechanical behavior of sliding stacks and the registry changes under bending. Most studies of registry changes have been for multilayer graphene with flat geometries \cite{KC_LJ_comp, Gao_PRL, Wijk_PRB, Reguzzoni_JPC}. 

Here we perform molecular dynamics simulations to show that under peeling-induced bending the behavior of multilayer graphene stack is complicated, yet understandable by registry effects. It turns out that peeling-induced bending generates complex sliding patterns that alter the local stacking and registry of multilayer graphene. This suggests the possibility to modify or even control the electronic properties of multilayer graphene by peeling or by creating localized kinks.  

%We make simulation setup similar to the experiment carried out in \cite{Mos} (ensimmaisen kuvan perusteella..??). We find out that the alingments of the layers change radically when the stack is bent and that this behavior is way too complicated to be described by any simple model. The slidind occurs both perpendicular and parallel to the bend axis. The parallel sliding is attributed to the multilayer graphene layers sliding in the valleys of the corrugation potential while the perpendicular sliding is natural provited that interlayer distances remain constant and there is no streching within layers. It has been noted that the stacking of the layers affect strongly to the elastic modules of the multilayer graphene, especially to the interlayer shear modulus \cite{ElasticConstantDFTTeor}. Thus the alingment of the adjacent layers is supposed to play key role in the graphene bending phenomena also mechanically.

%We perform molecular dynamics simulations and use the registry-dependent interlayer potential \cite{KC_potential}, which is known to describe the corrugation energy of multilayer graphene way better than the Lennart-Jones interaction which underestimates the corrugation energy by an order of magnitude \cite{LJOrder}. 

\section{Simulations}

\begin{figure}[t!]
	\centering
	\includegraphics[width=.45\textwidth]{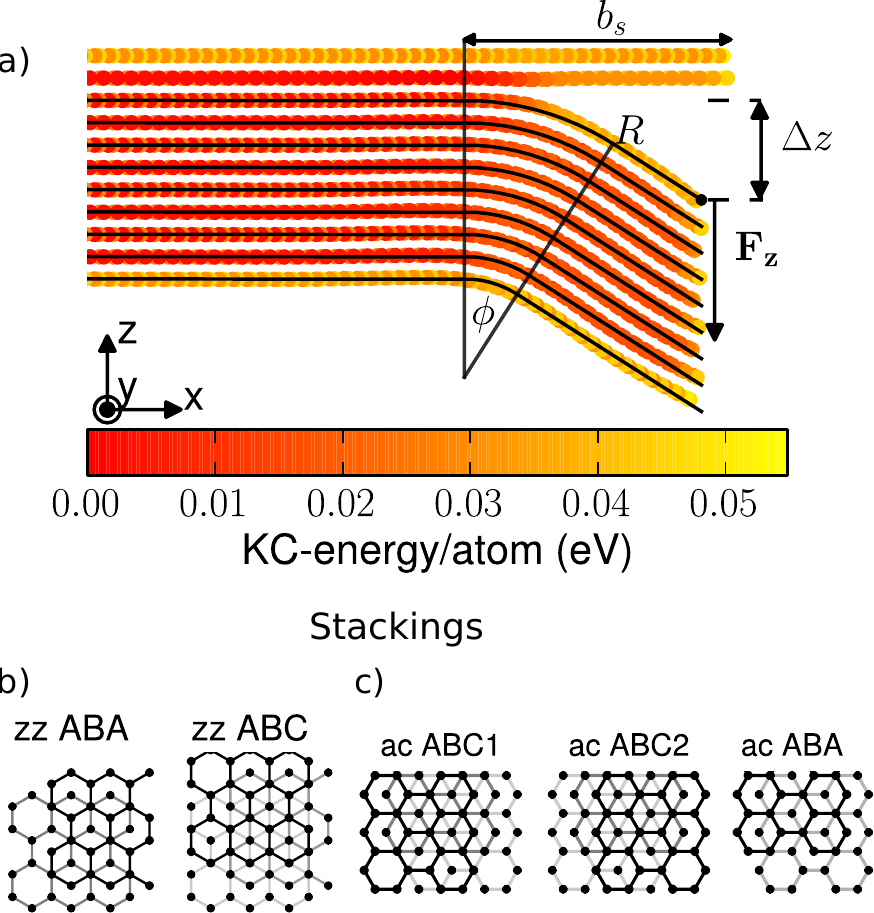}
	\caption{a) N=9 zigzag ABA-stacked multilayer graphene. The stack is peeled by pulling down the carbon atom marked by the arrow at constant velocity. The symbols and the solid lines fitted to individual bent layers give a simplified description of the large-scale geometry, as described in the text. Color represents the KC-energies of single atoms. \cmtr{b) Stackings of zigzag systems. c) Stackings of armchair systems.}}
	\label{pic:general}	
\end{figure}
% GEOMETRY

Fig.~\ref{pic:general} presents an example of a studied system. An N-layer multilayer graphene stack was peeled by \cmtr{displacing downwards the rightmost carbon atoms on top of the peeled stack, while letting them move freely in the $xy-$plane.} This simulation setup was designed to mimic recent experiments that used the tip of a nanoprobe to peel multilayer molybdenum disulphide \cite{Mos}. The upmost, unpeeled layer was fixed. In this peeling process the multilayer stack got inevitably bent. The systems were periodic in $\hat{y}$-direction \cmtr{with a constant unit cell size that fixed the width of the stack. All the free graphene edges were passivated by hydrogen.} Bending induced by peeling guaranteed sliding in which layers below always slid towards the peeled end relative to layers above. The systems were fixed at the left end but their lengths ($\approx 140$ \AA) made this constraint irrelevant. \cmtr{In this respect our simulation differed from previous ones that, due to tighter end constraints, resulted in puckering and bending-induced delamination instead of sliding\cite{Koskinen_CM, NIkiforov_prl}}. 

\begin{figure}[t!]
	\centering
	\includegraphics[width=.45\textwidth]{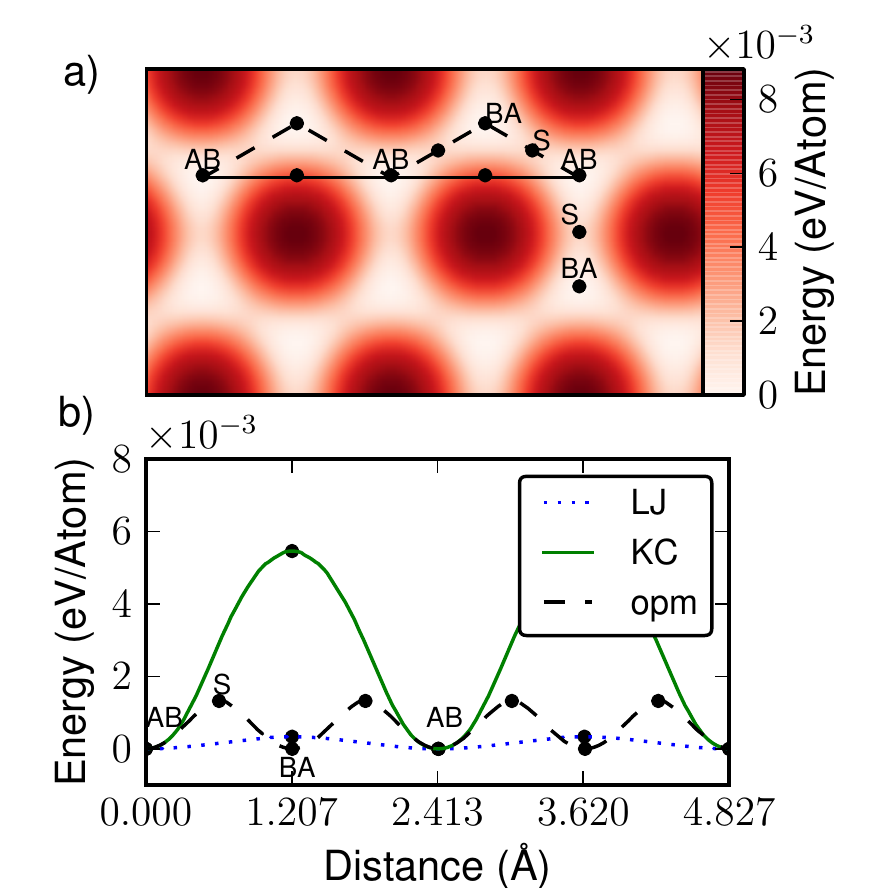}
	\caption{ Energetics for sliding in zigzag direction. a) Corrugation potential surface for graphene sliding on graphene (corrugation per surface area is obtained by dividing by $2.62$~\AA$^2$). b) Corrugation potential along the zigzag direction for KC- and LJ-potentials (solid line in a-panel) and the corrugation potential along the optimal staggered trajectory (dashed line in a-panel; only KC-potential).}
	\label{pic:KCLJ_corr_zz}	
\end{figure}

We considered zigzag (zz) systems where zigzag direction was parallel to $x$-axis and armchair (ac) systems where armchair direction was parallel to $x$-axis. In addition to this we considered also different stackings. For zigzag systems we considered two different stackings, ABA and ABC (Fig.~\ref{pic:general}b). For armchair systems we considered ABA-stacking and two inequivalent ABC stackings, ABC1 and ABC2 (Fig.~\ref{pic:general}c). While ABC2 is just ABC1 rotated around z-axis, it turned out that this rotation had significant effect regarding the sliding of the layers.

% METHOD
To perform molecular dynamics simulations, we used the LAMMPS package \cite{Lammps}. The intralayer carbon-carbon and carbon-hydrogen interactions were described by standard reactive bond order (REBO) potential \cite{REBO}. Because the in-plane Young's modulus of graphene is three orders of magnitude larger than the interlayer shear modulus \cite{ElasticConstantDFTTeor}, the mechanical behavior of multilayer graphene upon bending is dominated by interlayer shear. Hence it was important to have an interlayer potential that gives a fair description for the corrugation potential. In some studies \cite{InterlayerShearTeor, Nazem_JAP} modified Lennard-Jones (LJ) potential was used to describe the interlayer interactions. However, although the modification (\verb|epsilon_CC| = 45.44 meV instead of 2.84 meV) gives a reasonable corrugation potential, it gives unrealistically high interlayer adhesion energy. In our peeling simulations both corrugation and adhesion played important roles and had to be described with fair accuracy simultaneously. These requirements set us to use the registry-dependent interlayer potential by Kolmogorov and Crespi (KC)\cite{KC_potential}. \cmtr{The interlayer potential acts only between the nearest neighbors, which implies that prior to bending the different stacking geometries of Fig. \ref{pic:general}b and c are energetically equivalent.}

%The optimal sliding trajectory is marked by dashed line and we plot the corrugation potential along the solid and dashed lines in Fig.~\ref{pic:KCLJ_corr_zz}b.

\begin{figure}[t!]
	\centering
	\includegraphics[width=.45\textwidth]{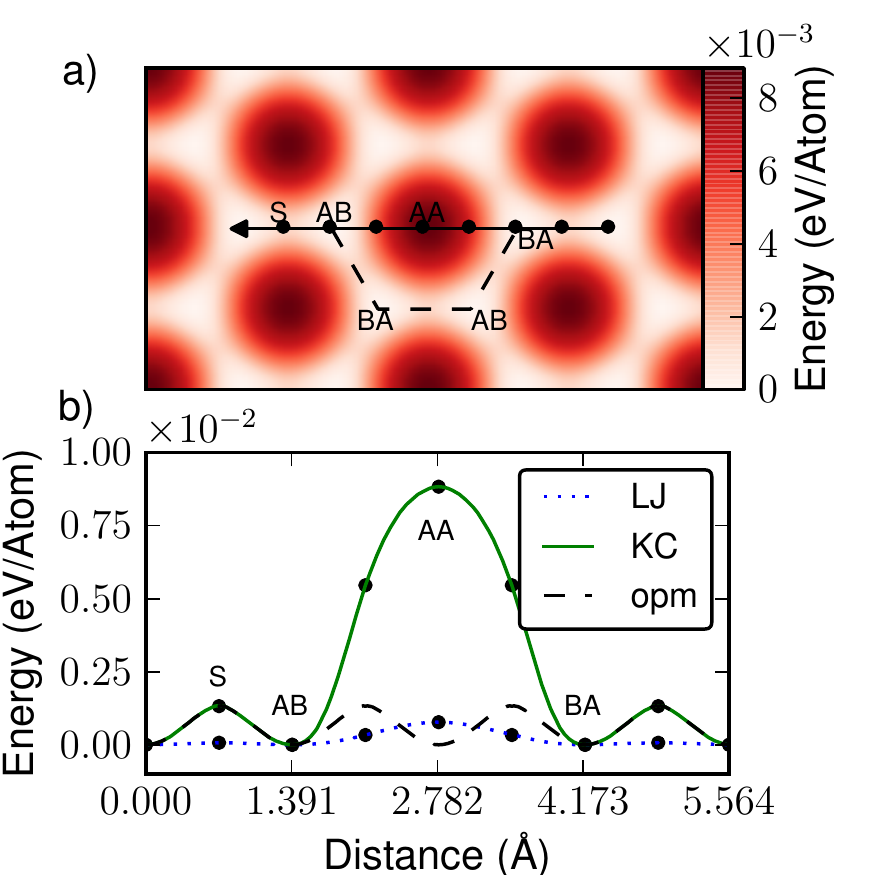}
	\caption{Energetics for sliding in armchair direction. a) Corrugation potential surface for graphene sliding on graphene. b) Corrugation potential along the armchair direction for KC- and LJ-potentials (solid line in a-panel) and the corrugation potential along the optimal staggered trajectory (dashed line in a-panel; only KC-potential).}
	\label{pic:KCLJ_corr_arm}	
\end{figure}

The corrugation potential surface given by the KC-potential is presented in Fig.~\ref{pic:KCLJ_corr_zz}a, together with visualization of sliding in zz-direction. For comparison also LJ-potential \cmtr{for the sliding part is shown in Fig. \ref{pic:KCLJ_corr_zz}b}. Sliding in armchair direction is visualized in Fig.~\ref{pic:KCLJ_corr_arm}. In both directions the optimal trajectory is not a straight line between two equivalent stackings but require staggered movement to avoid high corrugation barriers.

% SIMUL PROCESS
Simulations began by relaxation and thermalization of the system to $10$ K. We used Langevin dynamics with friction parameter of $2\times 10^{-4}$~fs$^{-1}$ and time step of $2$ fs. After thermalization we started to peel the system by pulling down the rightmost atoms on top of the peeled stack at a constant downward velocity of ${\Delta z}/{\Delta t} = 1$ \AA/ps, which turned out to be small enough for quasi-static dynamics. This constraint acted only in $-\hat{z}$-direction and the atoms were free to move in $xy$-plane. The peeling was continued for bending displacements up to $\Delta z = 30-40$ \AA. This displacement range turned out to be sufficient to display the most interesting new physics; continuing the peeling for larger $\Delta z$ did not bring essential new phenomena not already present at small $\Delta z$. 

% KC MOMENTUM
%We want to note the the KC-potential does not conserve momentum. $\nabla_i V(r_i, r_j) \neq -\nabla_j V(r_i, r_j)$ in general. Esimerkki?  

% LANGEVIN AND PRACTICE
    
%METHODS
% KC

We studied systems with thicknesses $\text{N}=4-10$. Zigzag and armchair were expected to behave differently during the peeling for two reasons: (i) The corrugation potential barrier in zigzag direction is lower and the required deviations from straight trajectories are smaller than for armchair direction. (ii) For zigzag the corrugation potential has only single minima per lattice period, whereas the corrugation potential for armchair system has two minimas per lattice period (Figs.~\ref{pic:KCLJ_corr_zz} and~\ref{pic:KCLJ_corr_arm}). The two minimas for armchair direction set layer pairs in different positions with respect to the anticipated sliding mechanisms.

%For armchair systems the BA-stacked pair has corrugation potential barrier much higher than AB-stacked pair when sliding to left (Fig.~\ref{pic:KCLJ_corr_arm}b). These properties related to system adge and stacking shall be considered thoroughly.
%

\section{Results}

\subsection{Large-scale geometry}

\begin{figure}[t]
	\centering
	\includegraphics[width=.4\textwidth]{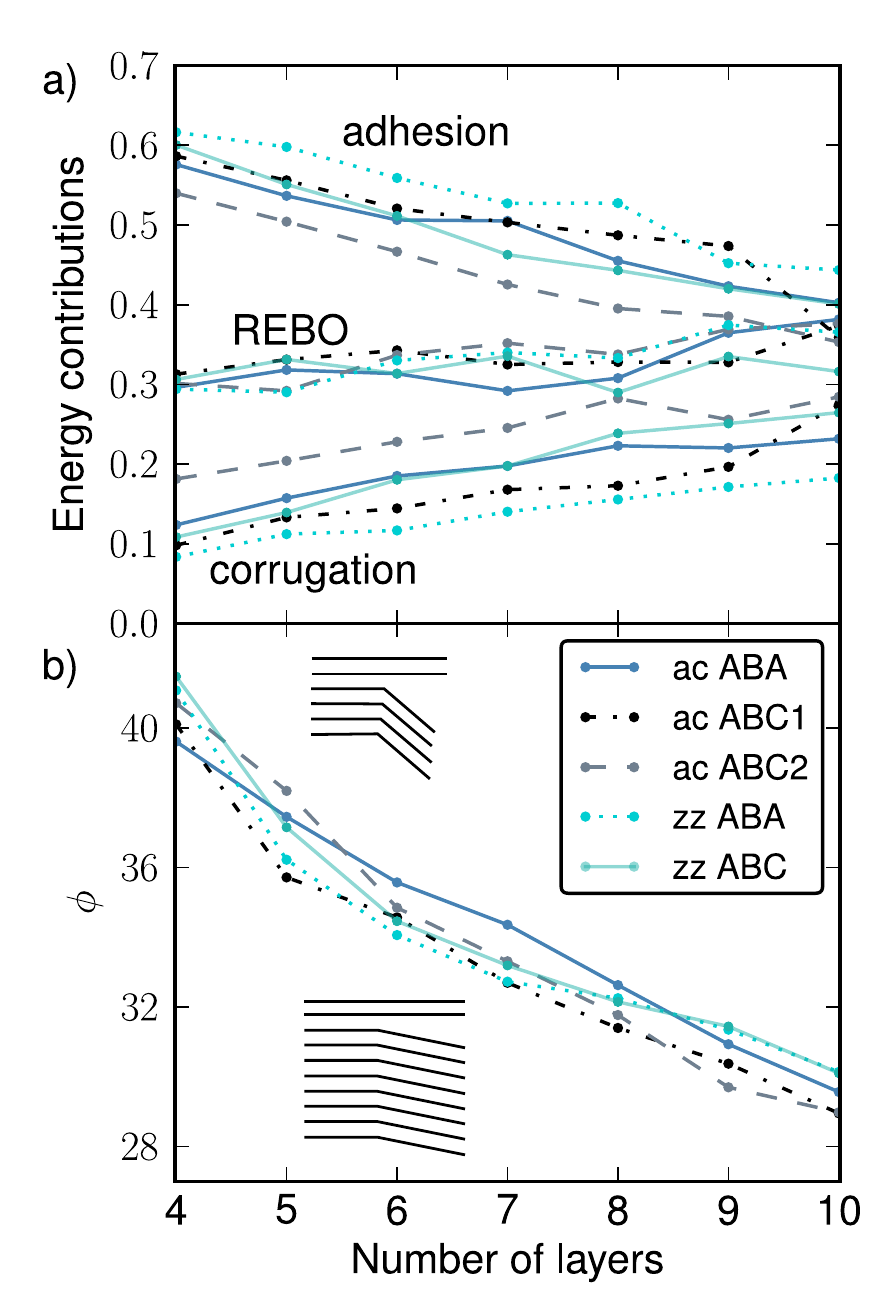}
	\caption{a) Average energy contributions (normalized to one) during the range  $\Delta z = 1 \ldots 30$~\AA. b) The average deflection angle $\phi$ during the range $\Delta z = 10 \ldots 20$~\AA. Insets illustrate the different behaviors of deflection angle for thin and thick stacks.}
	\label{pic:efracs}	
\end{figure}

Even though the corrugation energy plays important role in the bending process, its relative contribution to the total potential energy is small. In Fig.~\ref{pic:efracs}a we present the contributions of three different potential energies, the adhesion between the stack and the top layers (the cost due to peeling), the intralayer energy related to the bending and stretching of the layers (the REBO potential), and corrugation energy related to the registry of the graphene layers. The corrugation energy contributes the least of the three but increases in importance as the number of layers N increases. Also the importance of REBO energy, which consists mostly of bending energy of the individual graphene layers, increases as N increases. The adhesion energy is related to the peeled length $b_s$ of the stack and is independent of the number of layers. Hence its relative importance decreases as the number of layers increases. The importance of these energy contribution trends becomes clear when we consider the large-scale geometry of the stack.

To describe the geometry we introduce N curves, each of which consist of two straight lines connected by a circular arc. The description has three parameters: radius $R$ of the top layer, deflection angle $\phi$, and the deflection point $b_s$ (peeled length) (Fig.~\ref{pic:general}). The radii of other layers are derived from $R$ by requiring constant interlayer distances. At each instant this description was fitted to the atomic positions projected to $xz$-plane to give the deflection angle as a function bending displacement $\Delta z$. By taking the average of the deflection angle in the displacement range $\Delta z =  10 \ldots  20 \text{ \AA}$ for each system with different layer numbers N, we observe that the averaged deflection angle decreases when the number of layers increase (Fig.~\ref{pic:efracs}b). This is a natural consequence of the increasing relative importance of the contributions of corrugation and intralayer energies with respect to the N-independent peeling contribution.     

To illustrate this, consider given displacement $\Delta z$. For thin stacks the total energy is dominated by the adhesion energy which implies small peeling length and large deflection angle (Fig.~\ref{pic:efracs}a). For thicker stacks the intralayer and corrugation energies would be minimized by making the deflection angles $\phi$ small, but this would in turn increase the peeling lengths and increase the cost in adhesion energies. With increasing number of layers the relative contribution of the adhesion energy decreases and systems start to behave more in favor of REBO and corrugation energies, implying  smaller deflection angles (Fig.~\ref{pic:efracs}b).

\subsection{Bending-induced sliding}

\begin{figure}[t]
	\centering
	\includegraphics[width=.45\textwidth]{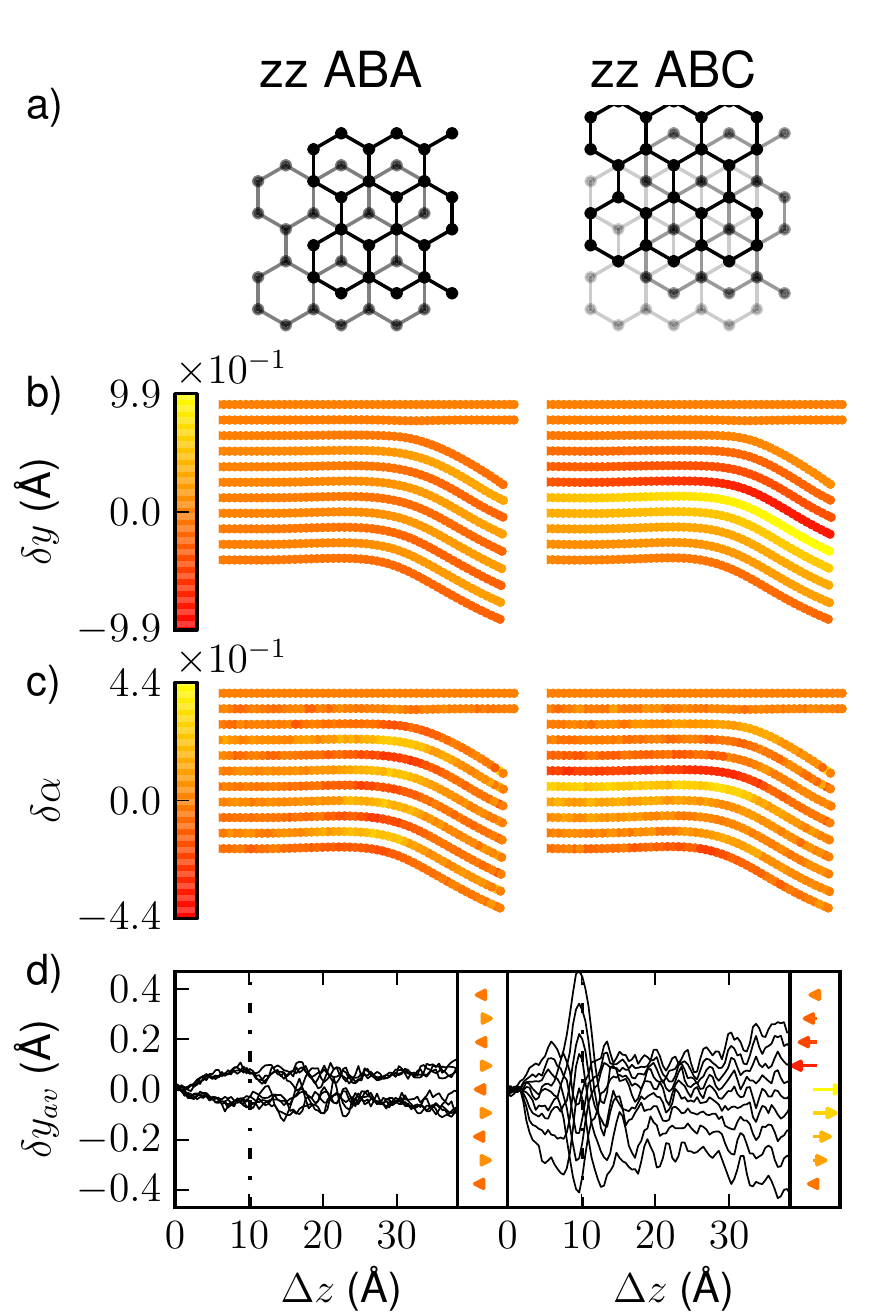}
	\caption{Sliding in zigzag systems. a) Stackings of the systems. b) Shifts in $\hat{y}$-direction. c) Average bond rotations. d) Average layer shifts as a function of $\Delta z$.}
	\label{pic:corror_zz}	
\end{figure}

\begin{figure}[t]
	\centering
	\includegraphics[width=.45\textwidth]{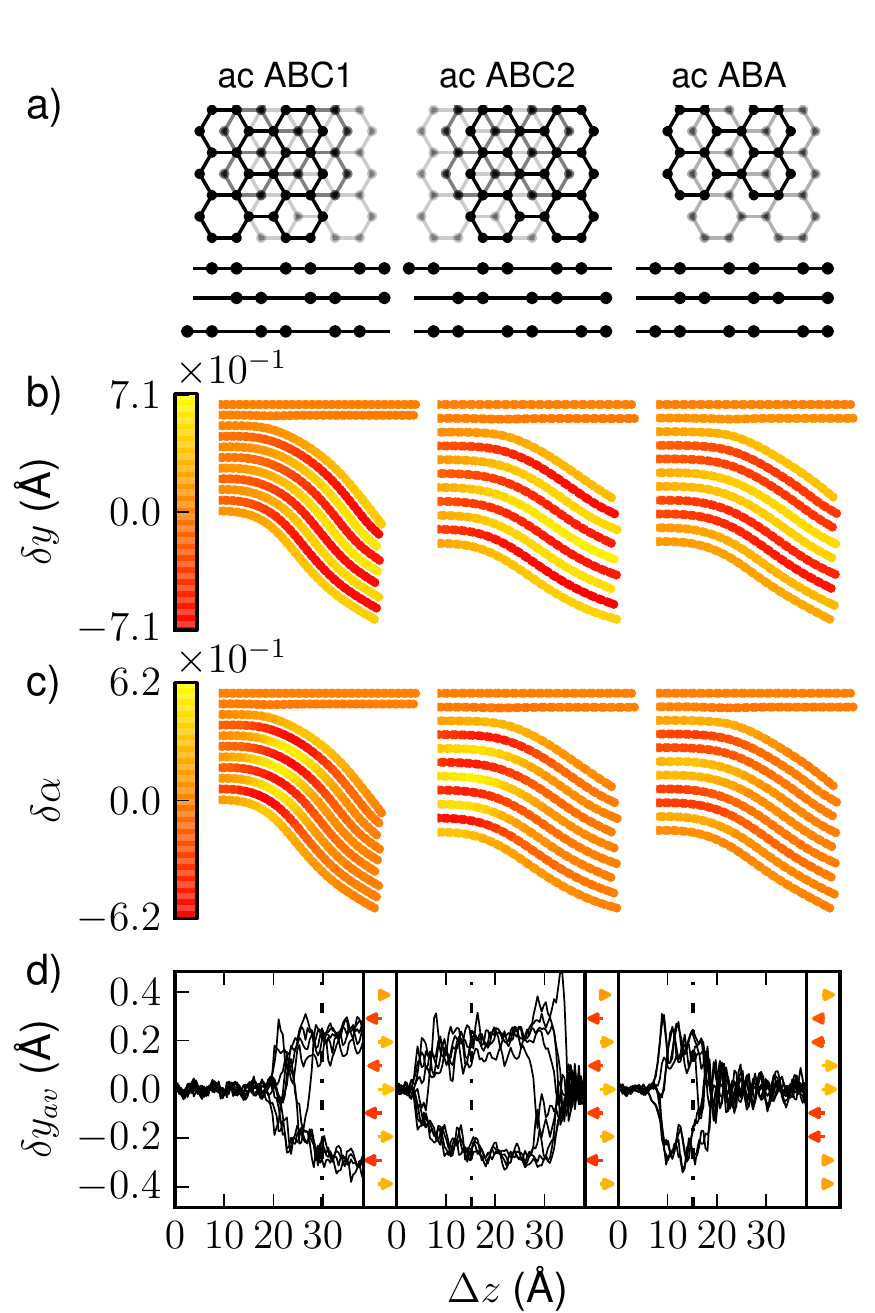}
	\caption{Sliding in armchair systems. a) Stackings of the systems. b) Shifts in $\hat{y}$-direction. c) Average bond rotations. d) Average layer shifts as function of $\Delta z$.}
	\label{pic:corror_arm}	
\end{figure}

As discussed earlier, upon bending the layers tend rather to slide than to compress due to the large Young's modulus of graphene. \cmtr{The observed maximum bond length variations were no more than $0.7$~\% and the streching and compression patterns of the stack showed similar behaviour to the sliding patterns in Figs. \ref{pic:corror_zz} and \ref{pic:corror_arm}.} The corrugation potential (Figs.~\ref{pic:KCLJ_corr_zz} and~\ref{pic:KCLJ_corr_arm}) suggested that sliding along straight line in $\hat{x}$-direction leads to high corrugation energies, especially in armchair systems. Let us now investigate in more detail the sliding behavior of the layers in the stack. 

%We find that the sliding occurs also in $\hat{y}$-direction for both zigzag and armchair systems, although the effect is much more pronounced in armchair systems. This sliding enables the layers to avoid the high corrugation energies related to AA-stacking even at the cost of small intralayer shear.

\subsubsection{Sliding in zigzag direction}

The optimal sliding trajectories in zigzag systems transforming from AB- to BA- to AB-stacking and the related $\hat{y}$-direction shifts are shown in Fig.~\ref{pic:KCLJ_corr_zz}a. In multilayer graphene the initial stacking defines the stacking for each layer pair. For ABA-stacked multilayer graphene every other layer is initially AB- and every other BA-stacked. As visible in Fig.~\ref{pic:KCLJ_corr_zz}a, the layer above tends to shift for AB-stacked pairs in $+\hat{y}$-direction and for BA-stacked in $-\hat{y}$-direction, because in this setting the layers above always move in $-\hat{x}$-direction relative to layers below. This results in layer-wise alternating sliding patterns. For ABA-stacked zigzag system every other layer shifts to $\hat{y}$-direction, and every other layer to $-\hat{y}$-direction (Fig.~\ref{pic:corror_zz}b). These alternating $\hat{y}$ shifts create intralayer shear into the graphene layers, which can be seen as bond rotations within kinks (Fig.~\ref{pic:corror_zz}c). Here we defined the average bond rotation as $\delta\alpha(\Delta z) = \frac{1}{3}\sum_{i=1}^3 \delta \alpha_{i}(\Delta z)$, where $\delta \alpha_{i}(\Delta z)$ are the rotations of nearest neighbors of atom $i$ around $z$-axis. The bond rotations vanish away from the kink, suggesting the absence of shear in the 'bulk' parts. The kink thus acts as a transition zone between two flat and regularly stacked multilayer graphene bulks without shear.

For ABC-stacked zz-system the situation is more involved. All layer pairs are initially AB-stacked, which means that all layers would like to shift in $\hat{y}$-direction when the bend displacement $\Delta z$ is small. However, for this reason the system accumulates considerable intralayer shear already in few layers thickness. When the shear towards $\hat{y}$-direction becomes too large, the tendency to avoid the high corrugation energy makes the layers to abruptly make a large shift in $-\hat{y}$-direction. This is visible in Fig.~\ref{pic:corror_zz}b, where the shift in $\hat{y}$-direction accumulates starting from the bottom until the fifth layer and the sixth then shifts in $-\hat{y}$-direction. ABC-stacked zigzag systems thus show an entire range of $\hat{y}$-shifts, unlike the other systems that show only definite shifts in opposite directions (Fig.~\ref{pic:corror_zz}d). 

\subsubsection{Sliding in armchair direction}

As already suggested by the corrugation potential, the systems with sliding in armchair direction are richer in behavior compared to systems with sliding in zigzag direction. To begin with, the difference in ABC1 and ABC2 stacking is highlighted in Fig.~\ref{pic:corror_arm}a. Upon bending the lower layers start to slide in $\hat{x}$-direction with respect to upper layers. Since for ABC1-stacking all layer pairs are initially AB-stacked, the corrugation barrier is small when the bending displacement $\Delta z$ is small and we therefore see no shifts in $\hat{y}$-direction until $\Delta z \approx 22$~\AA. When the shifts in $\hat{y}$-direction then finally occur, they become immediately large due to the tendency to avoid the large corrugation energy related to AA-stacking (Fig.~\ref{pic:corror_arm}). For ABC2-stacking, however, the situation is different. All layer pairs are initially BA-stacked, which means that without the $\hat{y}$-shift the AA-stacking would occur in early stages of bending and lead to large corrugation energy already at small $\Delta z$. As the slides between the layers are fairly short, the layers in ABC2-stacked armchair system remain sheared most of the simulation, excluding the very beginning and the very end (Fig.~\ref{pic:corror_arm}d).

%UUSIKSUUSIKS However, the systems can only avoid part of the corrugation energy as at some point the intralayer shear becomes as expensive as the bad registry and the energy reduction gained by shearing is lost. The ABC2-stacked system higlights these issues the most and it is separated from the other systems already when the different energy contributions are considered, the relative contribution of the corrugation energy is the largest for ABC2-stacked armchair system with all studied layer numbers (Fig.~\ref{pic:efracs}a).          

\begin{figure}[t]
	\centering
	\includegraphics[width=.45\textwidth]{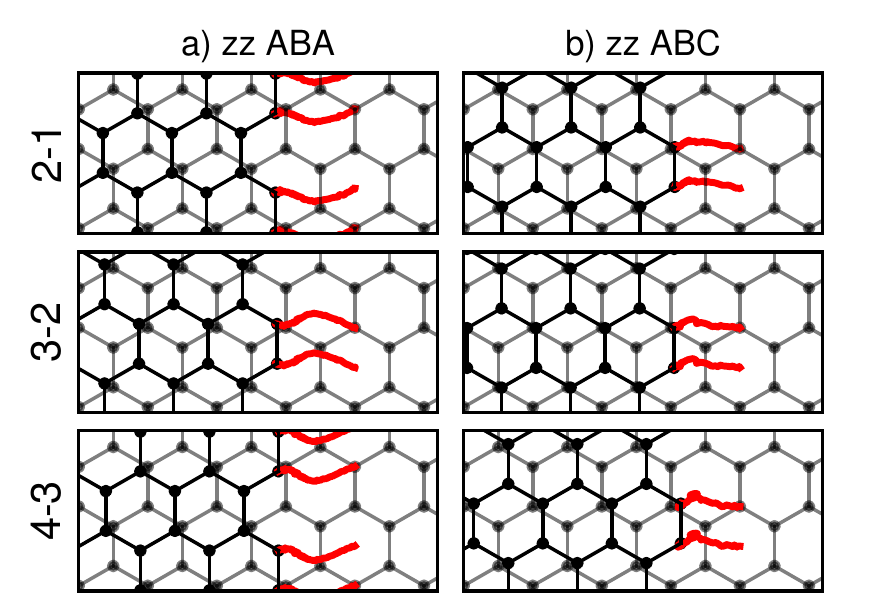}		
	\caption{Trajectories for atoms near the right edge of the zigzag system with N=9. The trajectories are presented relative to layer below. The layer pairs are listed starting from top and only few are shown for clarity. a) ABA-stacking. b) ABC-stacking.}
	\label{pic:paths_zz}	
\end{figure}

The ABA-stacked multilayer graphene is special in the sense that every other layer pair is AB- and every other BA-stacked. At small $\Delta z$ every other layer pair thus tends to create relative $\hat{y}$-shifts and every other layer pair tends to remain unshifted. This alternation manifests itself in a particular distribution of $\hat{y}$-shifts with two-layer pairs (inset Fig.~\ref{pic:corror_arm}d). The unshifted layer pairs are always AB-stacked, as for them there is no need for relative shifts at small $\Delta z$. However, the bottom layers of these AB-staked pairs are the top layers of the BA-stacked pairs below. For these BA-stacked pairs the shift is essential already in early stages of bending, as already discussed with the ABC2-stacked systems. Similarly the top layers of the AB-stacked pairs are the bottom layers of the BA-stacked pairs above, which again have to shift in $\hat{y}$-direction.

\subsubsection{Layer trajectories during peeling-induced sliding}

\begin{figure}[t]
	\centering
	\includegraphics[width=.45\textwidth]{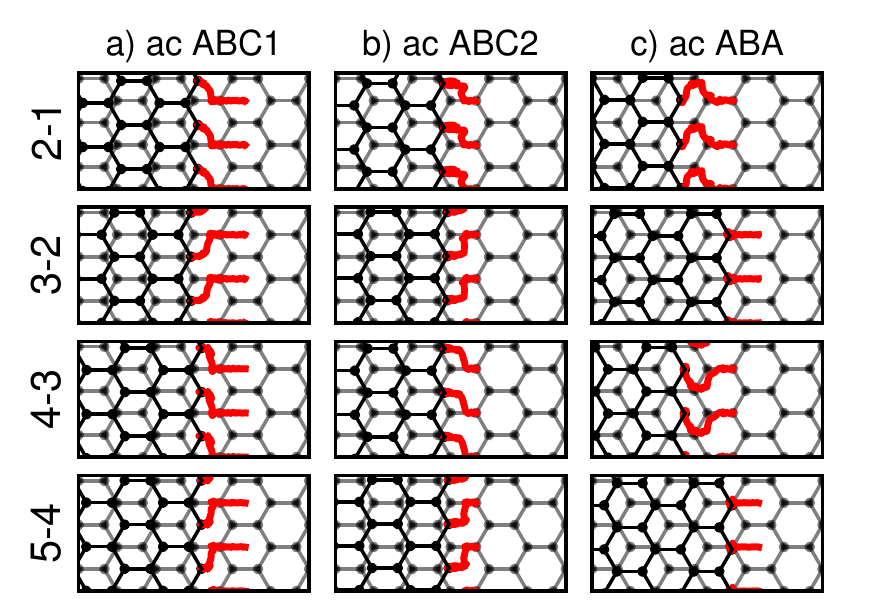}		
	\caption{Trajectories for atoms near the right edge of the armchair system with N=9. The trajectories are presented relative to layer below. The layer pairs are listed starting from top and only few are shown for clarity. a) ABC1-stacking. b) ABC2-stacking. c) ABA-stacking.}
	\label{pic:paths_arm}	
\end{figure}

To further illustrate the behavior of the sliding patterns of the layers, we visualize the trajectories of selected atoms near the right end of the stack (Figs.~\ref{pic:paths_zz} and~\ref{pic:paths_arm}). These trajectories represent well the 'bulk' graphite of the peeled part of the stack, and by following them one gets a picture of how the bulk part behaves upon bending. The trajectory is visualized by considering two atoms from a given layer pair and projecting their positions from upper layer to lower layer using the surface normal of the lower layer. After the initial stages of the simulations the layer pairs near the right end were parallel and straight which made this method justified. For clarity we do not plot atoms in upper layer that are to the right from the selected ones.

%For clarity we plot only atoms to the left from the chosen ones.

Trajectories show that especially for zigzag system with ABA-stacking the sliding occurs via alternation of stackings between AB to BA to AB (Fig.~\ref{pic:paths_zz}a). This is expected because for optimal trajectory the required intralayer shear is small. For zigzag systems with ABC-stacking the shifts in top layers are all in the same direction, which produces large intralayer shear since the shifts accumulate. Due to higher shears within the layers, the $\hat{y}$-shifts are not that pronounced although the trend of avoiding the AA-stacking is still evident (Fig.~\ref{pic:paths_zz}b).

Armchair systems, in contrast, always require large shears, but the onset of shifting depends on $\Delta z$ and on stacking. For ABC1-stacking we see relatively prolonged smooth sliding in $-\hat{x}$-direction followed by sudden shifts in $\hat{y}$-direction caused by the tendency to avoid the AA-stacking (Fig.~\ref{pic:paths_arm}a). This behavior is different from ABC2-stacking where the sudden shifts in $\hat{y}$-direction occur in the beginning of the trajectory (Fig.~\ref{pic:paths_arm}b). Also the ABA-stacked systems display their distinct behavior in the layer trajectories. The AB-stacked layer pairs do not have relative $\hat{y}$-shifts in contrast to BA-stacked layer pairs that avoid the AA-stacking by large $\hat{y}$-shifts (Fig.~\ref{pic:paths_arm}c).

These sliding patterns and $\hat{y}$-shifts are observed in all systems with different stackings, edges and thicknesses. The corrugation potential thus plays a central role in the sliding behavior of the layers when multilayer graphene gets bent upon peeling. The corrugation not only opposes the bending but it also forces the layers to shear within kink region in order to maintain favorable registry in the bulk part.

\subsection{Peeling force}

\begin{figure}[t!]
	\centering
	\includegraphics[width=.45\textwidth]{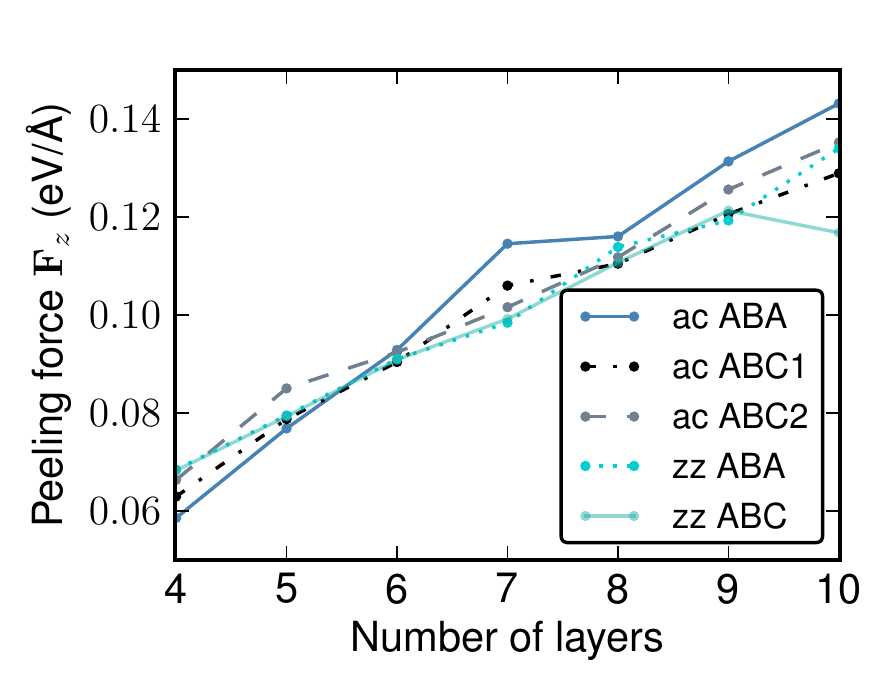}
	\caption{Averaged peeling forces for all systems, averaged between $\Delta z = 10 \ldots 20$~\AA.}
	\label{pic:forces}	
\end{figure}

The peeling force was not constant, as we pulled by small constant velocity. However, it turned out that the force oscillated around its average value during the entire simulation. Thus, to characterize the peeling force for each system we simply adopt the average force over the entire simulation. Although there were large differences in the behavior of different systems during the simulation, the differences averaged out to yield almost equal peeling forces for different systems with the same thicknesses (Fig.~\ref{pic:forces}). Moreover, the peeling force turns out to be roughly proportional to the number of peeled layers N. The N-dependence arises due to the corrugation and bending contributions of energy. \cmtr{Note that this direct proportionality is simple in comparison to the thickness-dependent frictional characteristics of multilayer graphene predicted for and observed in different types of experimental settings.\cite{Filleter2009,Cahangirov2013}} Adhesion does not explicitly depend on N, although a small implicit dependence does arise through the varying deflection angle.

We remark that pulling force was always necessary to keep the stack peeled and bent. We made several simulations where the force was released at different stages of the peeling process: the result was always restoring of the original completely flat multilayer graphene. This remained the case even if the two flat layers above were removed, that is, even if no adhesion contribution was present in the simulation. This is unlike the case of molybdenum disulphide where stable, or at least metastable, kinks could be made by bending the multilayer stack with a nanoprobe \cite{Mos}. This difference, however, is consistent with the much larger corrugation energy of MoS$_2$ layers \cite{Wang_mos, Cahan_mos}. Furthermore, to our knowledge there are no experimental observations of stable kinks in multilayer graphene, in agreement with these simulations.

\section{Conclusions}

We studied the peeling mechanics of multilayer graphene stacks using molecular dynamics with registry-dependent interlayer potential. We found that the corrugation energy plays major role in the creation of intricate sliding patterns and also in the generation of intralayer shears. The sliding patterns depend on both the initial stackings and the bending directions. These effects are particularly pronounced for systems where armchair direction is perpendicular to the bending axis. In fact, simulations suggest that different stackings could even be identified by their behavior upon bending. Because the stacking of multilayer graphene is known to affect its electronic and optical properties, the peeling and the concomitant bending also directly modify these properties.  

Although the different systems showed variations in the microscopic behavior, their large time- and length-scale quantities were fairly similar. This was seen in the peeling forces which were roughly the same for all systems of the same thickness. Our simulations used minimal periodic unit cells in $\hat{y}$-direction, which however was justified due to the high Young's modulus of graphene. This view is supported be related studies that have shown explicitly that a wider unit cell in $\hat{y}$-direction does not give raise to relevant additional effects \cite{Koskinen_CM}. 

%Our results give ideas how the stacking of multilayer graphene is modified by bending. We observe very rich and complicated behavior of the sliding patterns depending on many features including the initial stacking and the bending direction. 

In experiments multilayer graphene is not always nicely stacked and hence predicting sliding patterns in practice is difficult \cite{Pierucci_acs}. Moreover, we considered only cases where bending was parallel either to armchair- or to zigzag-direction. If bending should occur in other directions, it would require larger simulation cells and analysis of potentially even more complex sliding patterns; we let such systems to be the focus of future work. 

Provided that N-layer stack was thick enough (N$\gtrsim 3$) we found that the peeling is driven by the corrugation energy because the role of the lost adhesion energy gets diminished. We did not observe a single system where the kink would have been stable after releasing the peeling force, a situation different from multilayer molybdenum disulphide for which such stable kinks have been observed experimentally \cite{Mos}. This, however, is in line with the larger corrugation energy of MoS$_2$ compared to multilayer graphene. Nevertheless, the simulations suggest that the peeling process itself may transform the original stacking order of multilayer graphene and thus profoundly affect their electronic and optical properties.

%\bibliographystyle{apsrev4-1}
%\bibliography{BS_bib}{}
%\bibliographystyle{plain}


\end{document}